\documentclass[letterpaper, 10 pt, conference]{ieeeconf}  

\IEEEoverridecommandlockouts                              
\usepackage{graphicx} 
\usepackage{amsmath}
\usepackage{amssymb} 
\usepackage{fontenc}
\renewcommand{\t}{^{\mbox{\tiny\sf T}}}
\newcommand{\R}{\mathbb{R}}
\newcommand{\Q}{\mathcal{Q}}
\newcommand{\D}{\mathcal{D}}
\newcommand{\N}{\mathcal{N}}
\newcommand{\beq}{\begin{equation}\begin{array}}
\newcommand{\eeq}{\end{array}\end{equation}}
\renewcommand{\bmatrix}{\left[\begin{matrix}}
\newcommand{\ematrix}{\end{matrix}\right]}

\newtheorem{thm}{Theorem}
\newtheorem{Remark}{Remark}
\newtheorem{lem}{Lemma}

\title{\LARGE \bf
Nanometer Scanning with Micrometer Sensing: Beating Quantization Constraints in Lissajous Trajectory Tracking
}

\author{Matheus Lohse, Rafael S. Castro, Aur\'elio T. Salton and Minyue Fu
\thanks{This study was partially supported by CAPES, Brazil - Finance Code~001.}
\thanks{M. Lohse and A. Salton are with the School of Engineering, Federal University of Rio Grade do Sul, Porto Alegre, Brazil. E-mail: {\tt\footnotesize \{matheus.lohse,aurelio.salton\}@ufrgs.br}}%
\thanks{R.S. Castro is with the School of Technology, Pontifical Catholic University of
Rio Grande do Sul, Porto Alegre, RS, Brazil.
E-mail: {\tt\footnotesize rafael.castro@pucrs.br}}       
\thanks{M. Fu is with the School of System Design and Intelligent Manufacturing, Southern University of Science and Technology, Shenzhen, 518055, China. E-mail: {\tt\footnotesize fumy@sustech.edu.cn}}
}

\begin{document}

\maketitle
\thispagestyle{empty}
\pagestyle{empty}

\begin{abstract} This paper addresses the task of tracking Lissajous trajectories in the presence of quantized positioning sensors. To do so, theoretical results on tracking of continuous time periodic signals in the presence of output quantization are provided. With these results in hand, the application to Lissajous tracking is explored. The method proposed relies on the internal model principle and dispenses perfect knowledge of the system equations. Numerical results show that an arbitrary small scanning resolution is achievable despite large sensor quantization intervals.
\end{abstract}


\section{Introduction}


While the effects of quantization in feedback systems have been been under
investigation since the early 1960s \cite{Widrow1961}, our community has given it considerably more attention after the advent of networked control systems \cite{Walsh2002}. The logarithmic quantizer \cite{Elia2001}, bounds on data-rate-limited feedback \cite{Nair2000}, and the sector bounded approach \cite{Coutinho2010} are all examples of significant advances provided in the first decade of this century. Recent studies carry on the research with estimation \cite{Rego2021}, consensus \cite{Choi2022} and switched control \cite{Papadopoulos2018} to cite a few (see \cite{Fu2024} for a tutorial on the matter).

In motion systems quantization plays a very present role in the form of digital encoders for position and velocity measurements. However, when it comes to nanometer precision, digital encoders are usually precluded by some form of analog sensor, such as piezoelectric self sensing \cite{Liseli2020} or capacitive sensors \cite{Yong2013}. As a consequence, ``large'' displacements in this field are limited to the range of dozens of micro-meters. There is a clear opportunity to significantly enlarge the area of actuation of nanometer positioning systems, such as atomic force microscopes, by the use of inexpensive digital encoders, provided the effects of output quantization are mitigated.

The purpose of this paper is twofold: firstly, we would like to present a continuous time version of our previous result that dealt with the tracking of periodic references for discrete time systems subject to output quantization; and, secondly, we would like to explore their effects in the practical and interesting application of Lissajous tracking. The first part of the paper, therefore, will present  continuous time hitherto unpublished results that show how to achieve tracking of a system whose output is subject to uniform quantization. The second part will venture on the challenging task of tracking Lissajous trajectories when subject to quantized feedback.

We have addressed the problem at hand within the context of macro-micro manipulators where we achieved a partial theoretical result supported by experimental observations on a linear actuator \cite{Salton2020,Salton2021}. The role of positive realness was not yet clear at the time, and it would only came to light in the discrete time version of our study, presented in \cite{Salton2022}. Here we will show that these results are extendable to continuous time systems in a similar form. In particular, Section~\ref{sec:theory} will present conditions on both the reference and the pair plant-controller in order to achieve asymptotic tracking. Not all reference signals may be tracked by a closed loop system whose feedback is given by its quantized output -- a simple example is that of a reference bounded inside a quantization region. It is possible, however, to track sinusoidal references under mild assumptions regarding their amplitudes. The exploitation of these signals within the context of Lissajous trajectory tracking will be discussed in Section~\ref{sec:LT}.

\begin{figure}
\caption{The challenging task of tracking a reference subject to output quantization: even if $e = r-y=0$ it is possible that $\tilde e = r- q(y)\neq 0$. }
\label{fig:closed_loop}
\end{figure}

\section{Artificial quantization}
\label{sec:theory}

The feedback loop of interest is depicted in Fig.~\ref{fig:closed_loop} where the system output $y\in\R$ is shown to pass through the uniform quantizer $q(\cdot)$ before it is available for feedback. One may suppose it refers to a positioning stage $G(s)$ under the controller $C(s)$ whose output is measured by a digital encoder. The task of the controller is to achieve $y(t)\to r(t)$ as $t\to \infty$ from the error signal $\tilde e:=r-q(y)$. Clearly a challenging task because $\tilde e=0$ does not imply $r=y$. 

Let us define $q(\cdot)$ as: 
\beq{c}\label{eq:qy}
q(z):=j\cdot \Delta, \ \forall z\in \mathcal{Q}_j,
\eeq where $j\in \mathbb{Z}$, $\Delta \in \mathbb{R}^+$ is called the quantization interval and 
\beq{c}
\mathcal{Q}_j(\Delta):=\left\{ z\in\mathbb{R} \ | \  (j-0.5)\Delta \leq z < (j+0.5)\Delta\right\}\nonumber
\eeq is called the quantization region.  Furthermore, the quantization error is defined as $\epsilon_q := q(z)-z$ and satisfies
\beq{ccl}\label{eq:bound}
|\epsilon_q|&\leq& \Delta/2.
\eeq

We have set to ourselves the task of  beating the quantization error bound \eqref{eq:bound}: given the feedback loop of Fig.~\ref{fig:closed_loop}, we would like to achieve a steady state tracking error with a lower bound smaller than $\Delta/2$ and, if possible, achieve $\epsilon_q=0$.

Since the actual value of the output $y$ inside $\mathcal{Q}_j$ is unknown, it is clearly not possible to track \textit{static} references inside $\mathcal{Q}_j$. Instead, we must limit our attention to signals that allow one to determine $y(t)$ from $q(y(t))$, such as continuous-time signals persistently crossing quantization regions. In  principle, at the time instants $t=t_c$ when the crossings happen from $\Q_j$ to $\Q_i$, the output is recoverable by $$y(t_c)=\left(\frac{j+i}{2}\right)\Delta.$$ 

This leads to the conjecture that it may be possible to track references $r(t)$ that persistently cross quantization regions, such as sinusoids. Indeed, if the task at hand was limited to tracking periodic references with no quantization, one would simply resort to the Internal Model Principle (IMP) and replicate the dynamic components of the reference in $H(s):=C(s)G(s)$. Thus, a persistent output $y(t)$ would be maintained even when the input to $C(s)$ is zero, i.e., $e(t)\equiv 0$. However, when subject to quantization, the signal fed to the controller is $r(t)-q(y(t))\not\equiv 0$ even if $e(t)\equiv 0$. In order to allow the IMP to work as idealized, one should feed $H(s)$ with a signal $\tilde e(t)$ that satisfies $\tilde e(t)\equiv 0$ when $e(t)\equiv 0$. To that matter, we propose the control loop depicted in Fig.~\ref{fig:prop_closed_loop} with,
\beq{c}\label{eq:eq}
\tilde e(t):=q(r(t))-q(y(t)).
\eeq  
It is now clear that, by tracking $q(r(t))$ instead of $r(t)$, it is at least possible that $\tilde e(t) \equiv 0$ when $e(t)\equiv0$. Which effectively means that the system might find an equilibrium at $e(t)=0$.

\begin{figure}
\caption{The addition of the artificial quantization in the reference path allows $\tilde e = r- q(y)\neq 0$ when $e = r-y=0$. }
\label{fig:prop_closed_loop}
\end{figure}

In what follows we will explore the feedback loop in Fig.~\ref{fig:prop_closed_loop} and provide sufficient conditions on $H(s)=C(s)G(s)$ that guarantee that $\tilde e(t)\to 0$ as $t\to \infty$. But first,  conditions on the reference $r(t)$ and the quantization level $\Delta$ will be given in order to guarantee that $\tilde e(t)\equiv 0 \Rightarrow e(t)\equiv 0$.

\subsection{The Reference Signal} \label{sec:references}
\label{sec:ref}

Consider the reference signal given by,
\begin{align}
r(t) &=  \delta_0a_0+\sum_{i=1}^m a_i\sin(\omega_i t+\theta_i), \label{eq:ref}
\end{align}
where $\delta_0=0$ or 1, $m\ge 1,$ $\omega_i>0$, $a_i\ne 0$ for all $i$, and suppose $\omega_i$ are such that $r(t)=r(t+T)$. Since the quantization function must provide sufficient information to allow $r(t)$ to be recovered from $q(r(t))$, let us limit ourselves to the case where $r(t)$ contains at least one sinusoid and its magnitude range is sufficiently large with respect to the quantization step size $\Delta$. In particular, $q(r(t))$ must have at least $2m+\delta_0$ transitions per period in order to guarantee that $r(t)$, which has $2m+\delta_0$ free parameters, may be recovered from it. 

In order to formalize this statement, consider the reference as,
\begin{equation} \label{eq:ref2}
    r(t) = \phi(t)\rho
\end{equation}
with,
\begin{equation*}
    \phi(t) = \bmatrix 1 \ \ \sin(\omega_1 t) \ \ \cos(\omega_1t) \ \dots  \ \sin(\omega_m t) \ \ \cos(\omega_m t) \ematrix,
\end{equation*}
$\rho = \bmatrix a_0 & a_{11} & a_{12} & \dots & a_{21} & a_{22} \ematrix\t,\ \mathbb{R}^{2m\times1}$, and let us introduce to the following lemma:

\begin{lem} [Reference]
    Consider $r(t)=r(t+T)$ in \eqref{eq:ref2} and define $\mathcal{M}$ as the matrix containing $\phi(t)$ for all the $p$ time instants $t_1,\ t_2,\ \dots,\ t_p \in [0, T)$ when $r(t)$ crosses a quantization level: 
    $$\mathcal{M} := \bmatrix \phi(t_1)\\ \phi(t_2) \\ \vdots \\ \phi(t_m)\ematrix \in \mathbb{R}^{p\times (2m\times1)}.$$ Assume $r(t)$ is such that $p\geq 2m+1$ by design and suppose that $\mathrm{rank}(\mathcal{M}) = (2m\times1)$, then $r(t)$ is unique in the sense that any other function  
    $$\bar r(t) = \phi(t) \bar \rho,$$ achieves $q(r(t))=q(\bar r(t)),\ \forall\ t$ only if $r(t)=\bar r(t)$.
\end{lem}

\textbf{Proof:} From $q(r(t))=q(\bar r(t))$ it follows that $\phi(t) \rho = \phi(t) \bar \rho$ and that $\phi(t)(\rho-\bar \rho)=0$. For all $p$ time instants $t_1,\ t_2,\ \dots,\ t_p \in [0, T)$ when $r(t)$ crosses a quantization level. Hence, we have that
$$\bmatrix \phi(t_1)\\ \phi(t_2) \\ \vdots \\ \phi(t_m)\ematrix(\rho-\bar \rho)  = \mathcal{M}(\rho-\bar \rho)=0.$$ Since $\mathrm{rank}(\mathcal{M}) = (2m\times1)$ by assumption, the above is only true if $(\rho-\bar \rho)=0$, which, in turn, implies $r(t)=\bar r(t)$. \hfill $\square$

The shape of the reference as described above will be fundamental to prove asymptotic convergence of the error, provided both $r(t)$ and $y(t)$ are in the form \eqref{eq:ref}. In this case, $q(y(t))=q(r(t))$ implies $y(t)=r(t)$ and $e(t)=0$. We may now focus on a control strategy that achieves $q(y(t))\to q(r(t))$ as $t\to\infty$.

\subsection{Assymptotic tracking}


By expanding the reference in \eqref{eq:ref} as:
\begin{align}
r_0(t) &= a_0, \\ 
r_i(t) &= a_i\cos(\omega_it+\theta_i), i=1,\ldots, m.
\end{align}
it is clear that, in order to satisfy the IMP, $H(s)$ must contain a pair of complex poles for each sinusoid present in $r(t)$, and an integrator if $\delta=1$. That is, given,
\beq{c}\label{eq:H}
H(s):= \frac{\N(s)}{\D(s)},
\eeq then,
\begin{align}
\mathcal{D}_0(s)&=s; \ \mathcal{D}_i(s) = (s^2+ \omega_i^2). \label{eq:den}
\end{align}

With some abuse of notation, we use the Laplace operator ``$s$'' as the derivative operator, i.e., $sf(t) = \dot f(t)$. Then, it is easy to verify that,
\begin{align*}
\mathcal{D}_0(s) r_0(t) &= 0,\\\mathcal{D}_i(s)r_i(t) &= 0,\ \ i=1,\ldots, m. 
\end{align*}

In fact, since $H(s)$ expresses the input output relation between $\tilde e(t)$ and $y(t)$, and since $\D(s)r(t)=0$, it follows that 
\begin{align}
\N(s) \tilde{e}(t) &= \D(s)y(t)\nonumber \\
&= \D(s)y(t)-\D(s)r(t)\nonumber \\
&= -\D(s)e(t).  \label{eq:err}
\end{align} In words: if the IMP is satisfied, the error dynamics may be expressed by equation \eqref{eq:err}. 
With the above, we are ready to find conditions on $C(s)$ and $G(s)$ in order to achieve asymptotic tracking.


\begin{thm}\label{thm:1}
Suppose that $r(t)$ satisfies the Reference Lemma and assume that $H(s)$:
\begin{enumerate}    
    \item[$i$.] contains a simple pole at the origin if $\delta_0=1$;
    \item[$ii$.] contains a simple pole pair at $s=\pm j\omega$ for $i=1,\dots,\ m$;
    \item[$iii$.] has all other poles in $\mathrm{Re}\{s\}<0$.
    \item[$iv$.] is positive real;
\end{enumerate}    
\item  Then $e(t)=r(t)-y(t)\rightarrow 0$ as $t\rightarrow \infty$.
\end{thm}

\textbf{Proof:}
Let $(A,B,C,D)$ be the minimal realization of $H(s)$ such that (\ref{eq:err}) has the following state-space representation:
\begin{align}
\dot x & = A x + B\tilde{e} \nonumber \\
e&= -Cx-D\tilde{e}. \label{eq:ss}
\end{align}
Since $H(s)$ is positive real, i.e., passive, there exists a real symmetric positive-definite matrix $P$ and a storage function $V = x\t Px\geq 0$ such that,
\beq{l}
\dot V\leq -\tilde e\cdot e.
\eeq
Furthermore, $\tilde e\cdot e\ge0$ because the quantization function is monotonic (i.e., $r-y\ge0\Rightarrow q(r)-q(y)\ge0$). This implies that, in steady state, $V(t)$ must be confined in the region where $\tilde e\cdot e= 0$, i.e., $e=0$ or $\tilde{e}=0$. Which is equivalent to $\tilde{e}=0$ because $e=0$ implies $\tilde{e}=0$.

In steady state, given the fact that $\tilde{e}=0$ and that there are marginally stable roots of $\D(s)$, the output $y(t)$ must be of the  form:
\begin{align*}
y(t)&= \delta_0 a_0 + \sum_{i=1}^m a_i \sin(\omega_it+\theta_i).
\end{align*}
Note that this is the same form as $r(t)$ and that, as stablished by the Reference Lemma, $\tilde{e} =q(r)-q(y)=0$ implies that $y=r$ in steady state. That is, the tracking error is such that $e(t)\rightarrow 0$ as $t\rightarrow \infty$. 
\hfill$\square$

\subsection{Design guidelines}

The previous section has shown that tracking under output quantization is possible if the reference is properly designed and if the the direct path $H(s)$ between $\tilde e$ and $y$ is positive real and satisfies the IMP. Here we will provide brief comments on how to best approach the task of designing $C(s)$.


In particular, one should keep in mind that the positive realness of $H(s)$, that is, Re$\{H(j\omega)\}\geq 0$, has a straightforward frequency domain interpretation that $-90^o\leq \angle H(j\omega) \leq +90^o$. Furthermore, since the sum of Positive Real (PR) systems is also PR, one should let $H(s)$ take the form,
\beq{rl}\label{eq:H_design}
H(s) =& \delta_0\frac{k_0}{s}+\sum_{i=1}^m \frac{k_is}{s^2+\omega^2} \\ \ \ \ \ \ \ \ &+ \sum_{i=m+1}^{m+p}\frac{k_i}{s+p_i} + \sum_{i=m+p+1}^{m+p+q}\frac{k_i(s+z_i)}{s^2+c_i s + d},
\eeq
with,
\begin{enumerate}
\item[$1.$] $k_i\geq 0$ for all $i=0,1,\dots,m+p+q$.
\item[$2.$] $p\geq0$; $p_i>0$ for all $i=m+1,\dots,m+p$.
\item[$3.$] $q\geq0$; for all $c_i>0,\ d_i>0$ and $0\leq z_i\leq c_i$ for all $i=m+p+1,\dots,m+p+q$.
\end{enumerate}

Clearly, the first $m$ terms are used to satisfy the IMP, and the remaining terms are used to accommodate the plant dynamics. It remains to choose the gains $k_i$, $i=1,\dots, m+p+q$, which may be done according to traditional frequency domain techniques, provided one works in the form \eqref{eq:H_design} given above.

\begin{Remark}
     We are well aware of the limitations imposed by the positive realness of $H(s)$, a constraint not satisfied by any system with relative degree two or more. However, this seems to be a sufficient, but not necessary condition, which motivates further investigation on relaxations to the above. This is part of our ongoing research.
\end{Remark}

\section{Application to Lissajous Tracking}
\label{sec:LT}

Let us now turn our attention to the problem of tracking the so called Lissajous trajectory by, e.g., an imaging scanner. These trajectories are achieved by setting different sine waves to the individual axes of a 2D positioning system:
\begin{align}\label{eq:refL}
    r_x(t) =& x_{0} + a_x \cos (\omega_{x} t), \\
    r_y(t) =& y_{0} + a_y \cos (\omega_{y} t),
\end{align}
where the amplitudes $a_x$ and $a_y$ represent the sides of a rectangle centered at ($x_0$, $y_0$) that encloses the region of interest. In order to achieve a periodic trajectory, $\omega_{x}/\omega_y$ must be a rational number. A common choice is to set
\begin{equation*}
    \frac{\omega_x}{\omega_y} =   \frac{2N}{2N-1},
\end{equation*}
for some positive integer $N$, from which frequencies $\omega_{x,y}$ may be computed to achieve the desired frame rate $f$ as follows:
\begin{equation}
    \begin{array}{rcl}
         \omega_x = \dfrac{N\cdot f}{\pi}, &  & \omega_y = \dfrac{(2N-1)\cdot f}{2\pi}. 
    \end{array}
\end{equation}

The maximum distance between two adjacent scan curves, usually called the ``scan resolution,'' is approximated by \cite{Yong2013}:
\begin{equation}\label{calc_h}
    h \approx \frac{\pi a_x a_y}{N \sqrt{a_x^2+a_y^2}}.
\end{equation}

Here we will show that it is possible to achieve a scan resolution significantly smaller than the interval of quantization of the positioning sensor. In other words, it is possible to achieve a nanometer scan resolution, with a micrometer sensor quantization interval.

Note that the trajectory given by \eqref{eq:refL} is clearly within the scope of the current paper since it requires the tracking of a constant and a sinusoid: each axis of the positioning system must track a reference equal to \eqref{eq:ref} with $m=1$ and $\theta=0$.

\begin{figure}
\vspace{.4cm}
\includegraphics[width=\columnwidth]{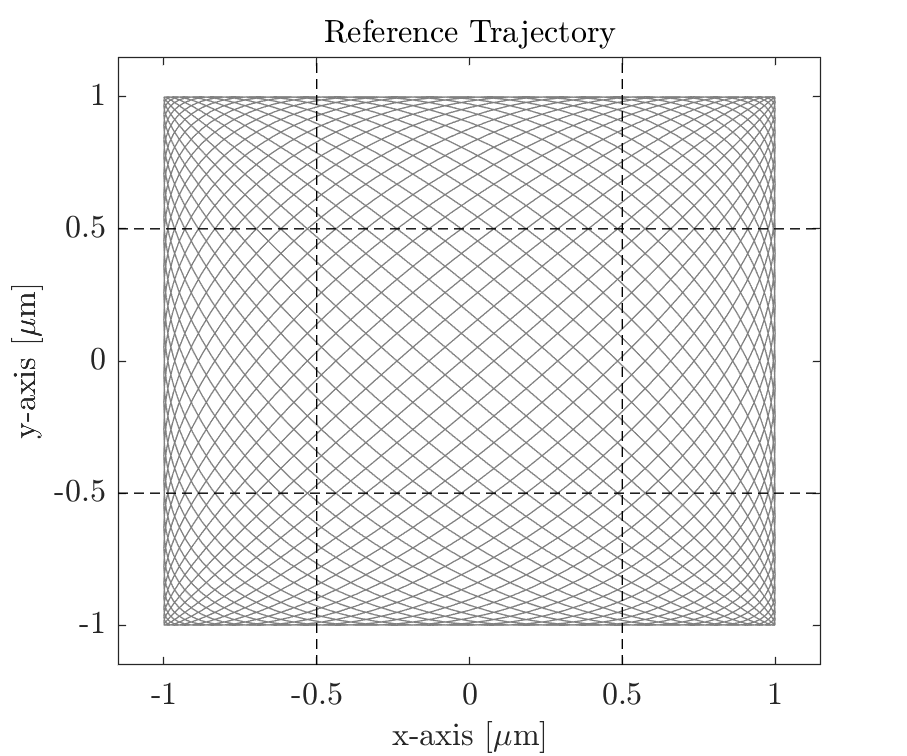}
\caption{A typical pattern of the Lissajous trajectory given by the gray line along with the quantization regions delimited by the dashed lines. }
\label{fig:lissajous}
\end{figure}

Let us assume that the 2D positioning system at hand is comprised of two decoupled axes, each formed by a mass-damper,
$$G_{x,y}(s)= \dfrac{Y_{x,y}(s)}{U_{x,y}(s)} = \dfrac{b}{s(s+a)},$$
and subject to a quantized sensor available for feedback. Since the plant itself has a pole at the origin, in order to track the trajectory defined by \eqref{eq:refL} the controllers must include a resonant mode,
\begin{equation}
    C_{x,y}(s) =  \frac{sk_{x,y}}{s^2+\omega^2_{x,y}},
\end{equation} 
where $\omega_{x,y}$ are given by the choice of trajectory, and $k_{x,y}$ are free parameters to be designed in order to guarantee closed loop stability and achieve the desired dynamic performance. 

A quantization interval of $\Delta = 1\ \mu$m is imposed to both feedback loops while we would like to track an area of $a_x\times a_y = 1\times 1\ (\mu\mathrm m)^2$. We have set $N=30$ and a frame rate of $f=1$ Hz, resulting in $\omega_{x} = 30\cdot 2\pi$ rad/s, $\omega_{y} = 29.5\cdot 2\pi$ rad/s, leading to the reference trajectory presented in Fig.~\ref{fig:lissajous}. We are effectively attempting to a scan resolution of $h\approx70$ nm while subject to a sensor resolution of $\Delta=1\mu$m.
\begin{figure}
\includegraphics[width=\columnwidth]{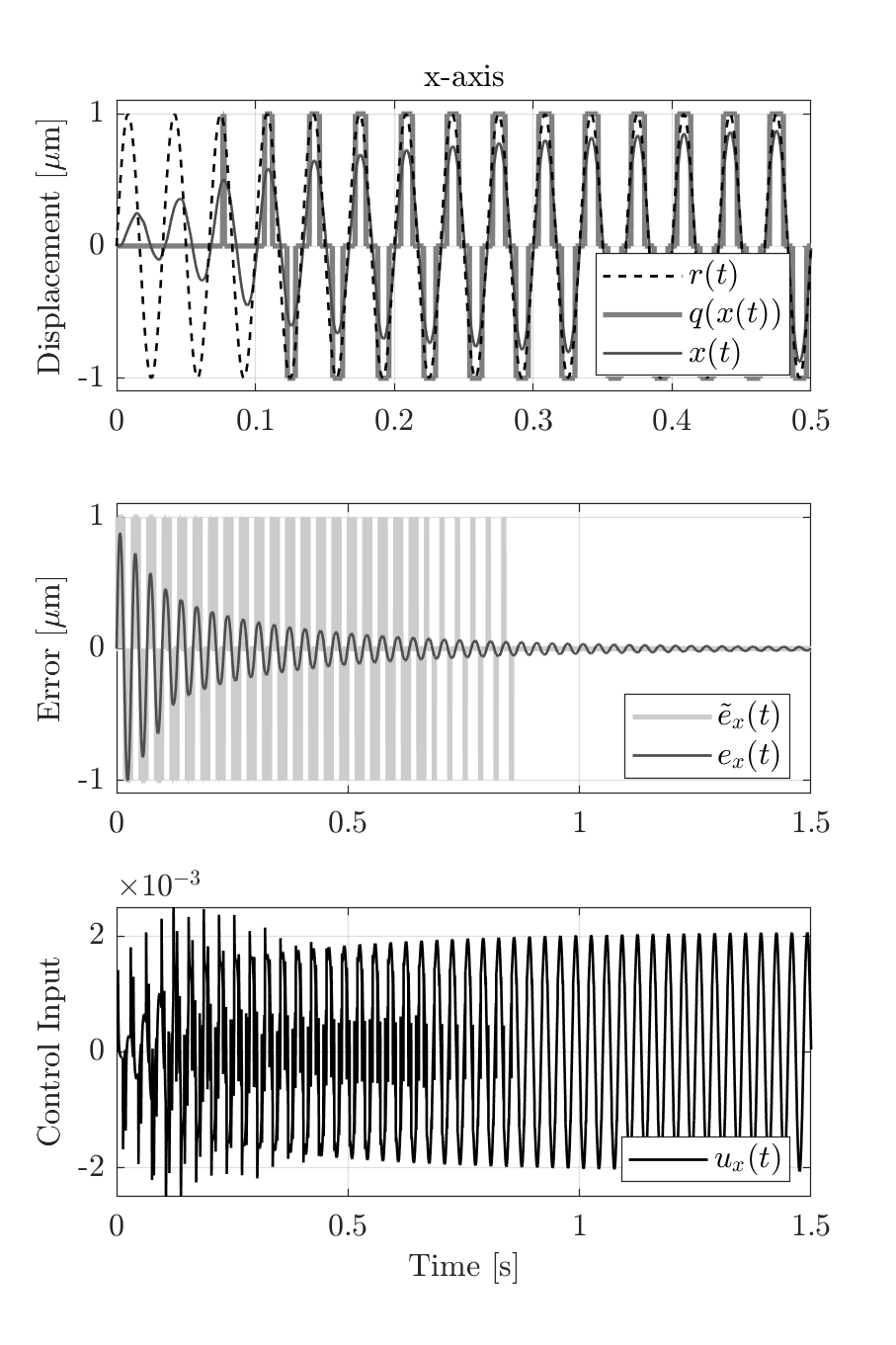}
\caption{Time evolution of the $\mathrm{x}$-axis in three plots. Top: displacement; Middle: errors; bottom: control effort.}
\label{fig:time_x}
\end{figure}

\begin{figure*}
\center
\includegraphics[scale=.44]{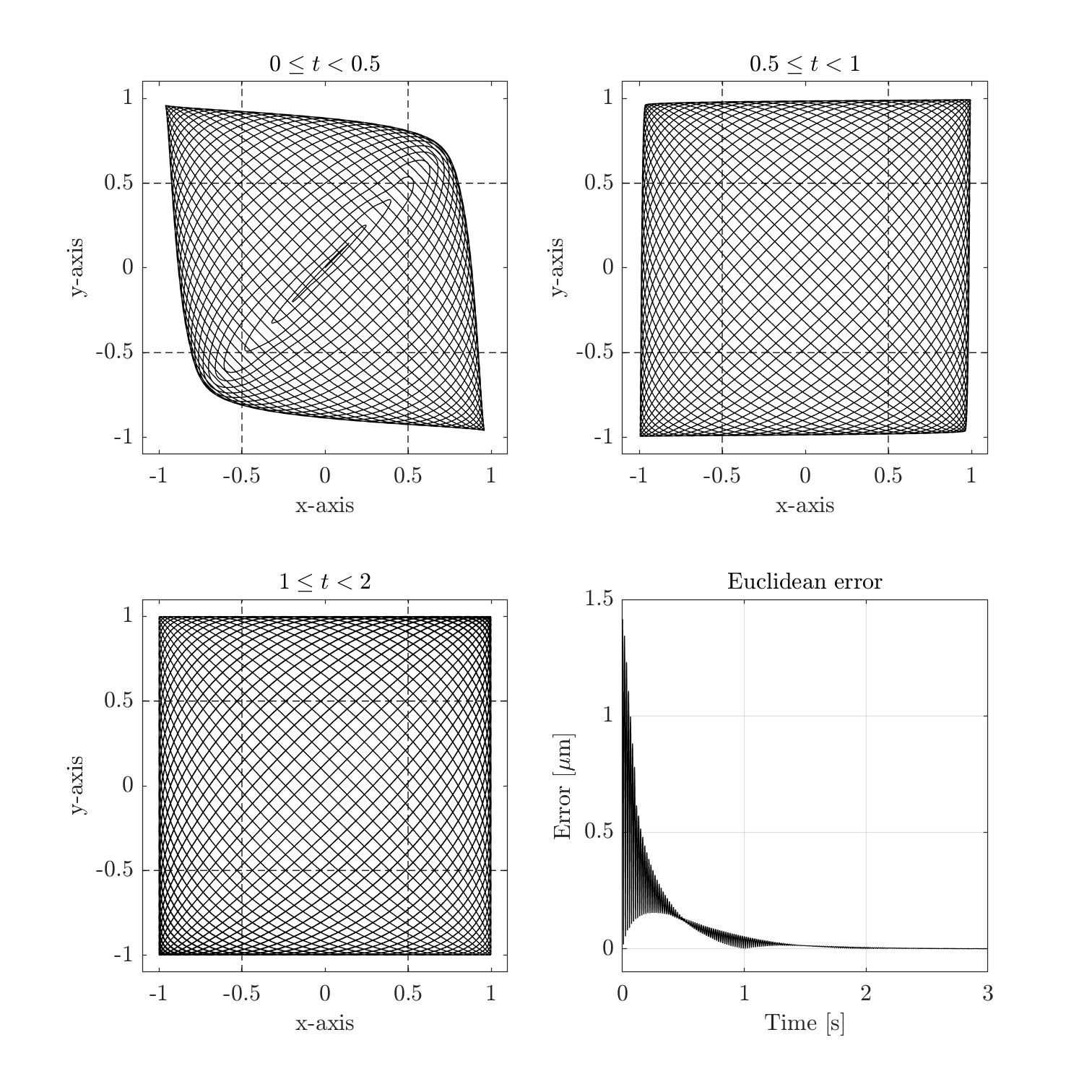}
\vspace{-1cm}
\caption{Time evolution of the overall lissajous trajectory. The bottom right plot shows the euclidean error of the trajectory asymptotically approaching zero.}
\label{fig:time_lissajous}
\end{figure*}

In order to simplify the presentation, we use the same system as in \cite{Salton2020} where $a=10$ and $b = 1.7\times 10^7$, and set $k_1=k_2=k_3=10$ in,
\begin{equation}
    H_{x,y}(s) = \frac{k_1}{s} + \frac{k_2s}{s^2+\omega_{x,y}} + \frac{k_3}{s+a}.
\end{equation}
This simple choice results in the following controllers:
\begin{align}
C_x(s) =& \frac{0.002s^3 + 0.012s^2 + 41.8s  + 209}{(s^2+\omega_x^2)(p^{-1}s+1)}\\
C_y(s) =& \frac{0.002s^3 + 0.012s^2+ 40.4s+  202}{(s^2+\omega_y^2)(p^{-1}s+1)}
\end{align} where $p>>\omega_{x,y}$ is used to adjust causality. This extra pole is necessary due to the fact the plant is a second order system and a noncausal controller would be required to achieve positive realness of $H(s)$. As mentioned before, this seems to be a sufficient, but not necessary condition to achieve asymptotic tracking. Simulation results presented from here on will show this is indeed the case.

Figure \ref{fig:time_x} shows the $\mathrm{x}$-axis of the positioning system tracking $r_x(t)$. The middle plot shows the interesting behavior of the error: after  $\tilde e(t)$ settles at the origin, the output is generated by the stable and marginally stable modes in $H(s)$, as the stable modes asymptotically decay to the origin, the trajectory converges to $e(t)=0$ in steady-state. The bottom plot shows the discontinuity of $\tilde e$ being reflected at the input signal in the form of large spikes. However, from the moment $\tilde e = 0$, the control signal enters its steady-state form, and presents no further spikes.

We will now combine both axes in order to achieve the desired Lissajous trajectory. These results are presented in Fig.~\ref{fig:time_lissajous} for different time intervals. Clearly, after approximately one second, the difference between the system trajectory and the reference output is imperceptible to the naked eye. To quantify the tracking performance we also show the Euclidean difference in the right bottom plot of the figure. After two seconds the trajectory error is below ten nanometers and after three seconds it is below one nanometer. This indicates that it is possible to achieve an even better tracking resolution if one chooses to adjust the parameters in \eqref{eq:refL}. In fact, by adjusting $N$ in \eqref{calc_h} it is possible to achieve an arbitrarily small scan resolution.

\begin{figure}
\includegraphics[width=\columnwidth]{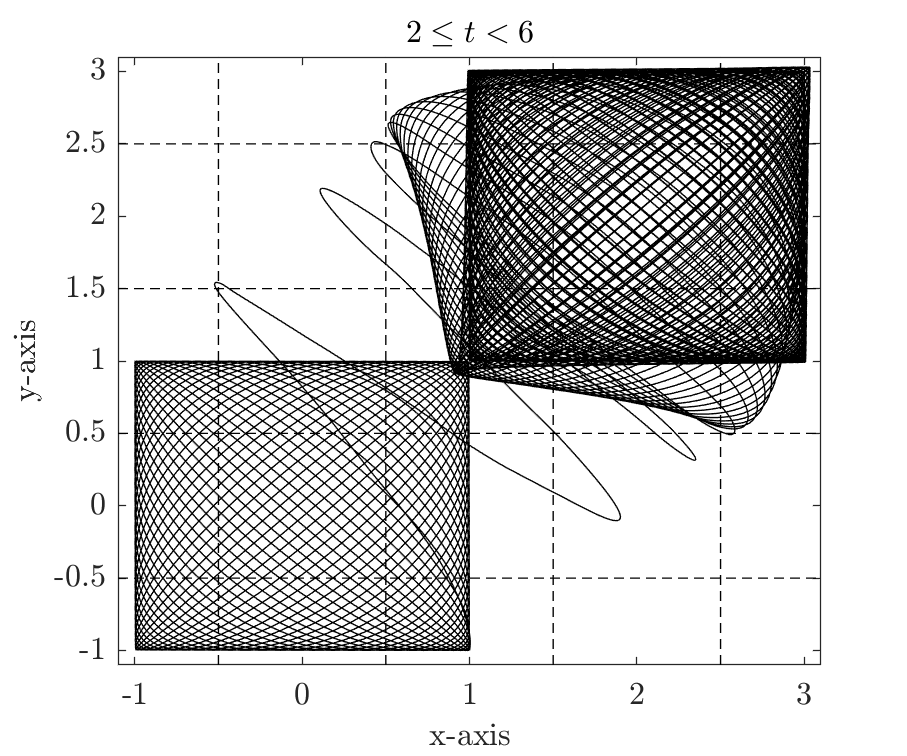}
\caption{At time $t=3$ s the system moves from a trajectory centered at the origin to one centered at $(x_0,y_0)=(2,2)$ illustrating that much larger scanning areas could be achieved.}
\label{fig:time_step}
\end{figure}

\begin{figure}
\includegraphics[width=\columnwidth]{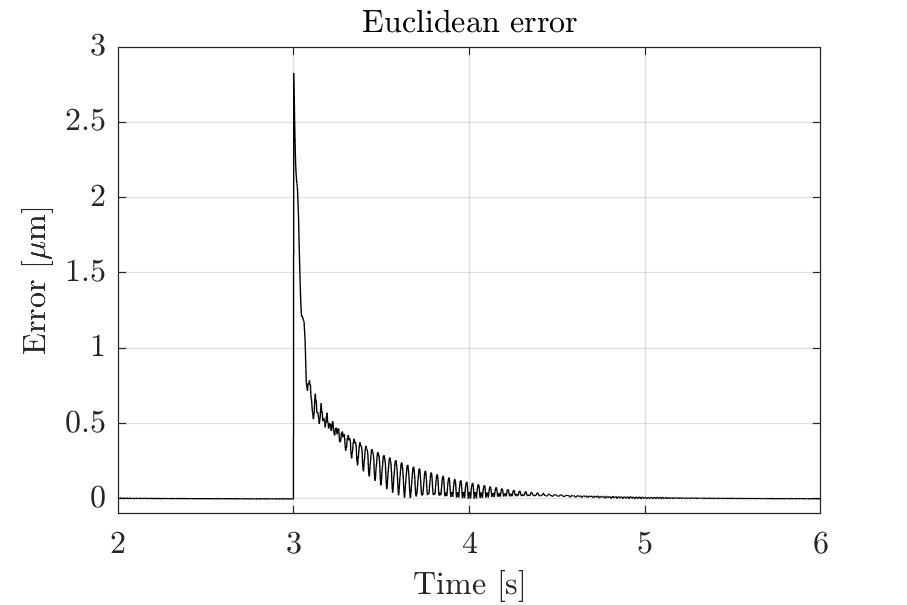}
\caption{The Euclidean error of the trajectory depicted in Fig.~\ref{fig:time_step}.}
\label{fig:time_step_euclidean}
\end{figure}

Our last simulation shows a step change being performed in the Lissajous trajectory (Fig.~\ref{fig:time_step}). At time $t=3$ s the system moves from the trajectory centered at the origin to one centered at $(x_0,y_0)=(2,2)$. Naturally, this step change could be significantly larger, indicating the possibility to cover large scanning areas. The corresponding Euclidean error is shown in Fig.~\ref{fig:time_step_euclidean}.

 \pagebreak
 
\section{CONCLUSION}

This paper has shown how to achieve asymptotic Lissajous tracking in the presence of output quantization. Theoretical and numerical results were provided to validate the central idea of the paper: the implementation of an artificial quantization function at the reference. While the results presented here could be extended to different trajectories such as the Rosetta and Cycloid, we are currently investigating forms of relaxing the passivity constraint.


\end{document}